\documentclass[aps,prc,twocolumn,showpacs]{revtex4}
\usepackage{amssymb}
\usepackage{amsmath}
\usepackage{graphicx}
\usepackage{lscape}
\usepackage{booktabs}
\usepackage{epsfig}
\begin{document}

\title{Higher moment singularities explored by the net proton non-statistical
fluctuations}

\author{Dai-Mei Zhou$^1$, Ayut Limphirat$^{2,5}$, Yu-liang Yan$^3$, Yun
Cheng$^1$, Yu-peng Yan$^{4,5}$, Xu Cai$^1$, Laszlo P. Csernai$^{6,7}$,
and Ben-Hao Sa$^{1,3,4}$}

\affiliation{ $^1$ Institute of Particle Physics, Central China Normal
University, 430082 Wuhan, China \\ and Key Laboratory of Quark and Lepton
Physics (CCNU), Ministry of Education, China.\\
$^2$ Department of Applied Physics, Faculty of Sciences and Liberal Arts,
Rajamangala University of Technology Isan, Nakhon Ratchasima 30000,Thailand. \\
$^3$ China Institute of Atomic Energy, P. O. Box 275 (10), 102413 Beijing,
China.\\
$^4$ School of Physics, Institute of Science, Suranaree University of
Technology, Nakhon Ratchasima 30000, Thailand.\\
$^5$ Thailand Center of Excellence in Physics (ThEP), Commission on Higher
Education, Bangkok 10400, Thailand. \\
$^6$ Department of Physics and Technology, University of Bergen, N-5007,
Bergen, Norway. \\
$^7$ Wigner Research Center for Physics, H-1525 Budapest, Pf. 49, Hungary.
}

\begin{abstract}
We use the non-statistical fluctuation instead of the full one to explore
the higher moment singularities of net proton event distributions in the
relativistic Au+Au collisions at $\sqrt{s_{NN}}$ from 11.5 to 200 GeV
calculated by the parton and hadron cascade model PACIAE. The PACIAE results
of mean ($M$), variance ($\sigma^2$), skewness ($S$), and kurtosis ($\kappa$)
are consistent with the corresponding STAR data. Non-statistical moments are
calculated as the difference between the moments derived from real events and
the ones from mixed events, which are constructed by combining particles
randomly selected from different real events. An evidence of singularity at
$\sqrt{s_{NN}}\sim$ 60 GeV is first seen in the energy dependent
non-statistical $S$ and $S\sigma$.
\end{abstract}

\date{\today}

\pacs{25.75.Dw, 24.60.Ky, 24.10.Lx}

\maketitle
%%%%%%%%%%%%%%%%%%%%%%%%%%%%%%%%%%%%%%%
\section {Introduction}
%%%%%%%%%%%%%%%%%%%%%%%%%%%%%%%%%%%%%%%
Describing the QCD phase diagram as a function of temperature $T$ and baryon
chemical potential $\mu_B$ is one of the fundamental goals of heavy-ion
collision experiments \cite{adams}. The finite temperature lattice QCD
calculation at $\mu_B$=0 predicts that a crossover transition from the
hadronic phase to Quark Gluon Plasma (QGP) phase may occur at temperature of
170-190 MeV \cite{aoki1,aoki2}. However, a QCD based model calculation
indicates that the transition could be first order at large $\mu_B$
\cite{bowman}. Once the location of the QCD critical point (QCP), where the
phase transition proceeds from first order to crossover, is  identified, the
global structure of the phase diagram is known \cite{steph,gavai}.

A characteristic feature of QCP is the divergence of the correlation
length $\xi$ \cite{star} and the extremely large critical fluctuations
\cite{steph1}. In a static and infinite medium, various moments of conserved
quantities such as the net-baryon, net-charge, and net-strangeness are
related to the correlation length $\xi$ \cite{koch}. Typically the variance
($\sigma^2$) of these distributions is related to $\xi$ as $\sigma^2\sim\xi^2$
\cite{steph1}. It is pointed out in \cite{step2} that higher moments of
conserved quantity event distributions, measuring deviations from a Gaussian,
are sensitive to the QCP fluctuations.

In the first quotation of \cite{cser} the hadronic and quark-gluon coexisting
phases in a small volume have been investigated with a simple effective model.
Assuming this finite system is in a heat reservoir, the order parameter
(energy density $e$) fluctuation near the phase transition is studied in a way
similar to the Landau theory. A positive skewness ($S$) was predicted and a
negative kurtosis ($\kappa$) was expected for the extensive thermodynamical
quantities. In the second quotation of \cite{cser} the dynamical change of
skewness and kurtosis was analyzed during hadronization of QGP. It was shown
that the skewness changes from negative to positive during the transition.
While the kurtosis is positive in the initial dominantly QGP phase and after
hadronization, but it is negative during the process of transition.
Similarly, it was predicted in the effective theory \cite{step2,asaka} that a
crossing of the phase boundary may result in a change of sign of skewness as
a function of the energy density. It was also reported recently that the sign
of kurtosis could be negative as well, if the QCP is approached from the
crossover side of the QCD phase transition \cite{step3}.

The products of higher moments, such as $S\sigma$ and $\kappa\sigma^2$, are
related to the ratio of conserved quantity number susceptibilities ($\chi$):
$S\sigma\sim\chi^{(3)}/\chi^{(2)}$
and $\kappa\sigma^2 \sim\chi^{(4)}/\chi^{(2)}$ \cite{sa}. It is predicted
that the conserved quantity event distribution becomes non-Gaussian and the
susceptibility diverges when QCP is approached. This causes $S\sigma$ and
$\kappa\sigma^2$ to change significantly.

Recently, QCP and the higher moments of conserved quantity event
distribution in heavy ion collisions at BNL Relativistic Heavy Ion
Collider (RHIC) energies have aroused further interest both
experimentally \cite{star,gupta,luo} and theoretically
\cite{rajiv,scha,pnjl,liuyx,dse,yang,peter,zhim}. On the
experimental side, a method to determine $T_c$ based on data has
been proposed in \cite{gupta}. By comparing the lattice results with
the RHIC BES (Beam Energy Scan) fluctuation data of variance,
skewness, and kurtosis systematically, the critical temperature is
determined to be $T_c=175^{+1} _{-7}$ MeV. However, no evidence of
singularity in energy dependence (at given centrality) and/or
centrality dependence (at given energy) has been reported. On the
other hand, copious models have been further proposed: such as the
Lattice QCD \cite{rajiv}, 2+1 flavor Quark-Meson model and
Polyakov-Quark-Meson model \cite{scha}, Polyakov-Nambu-Jona-Lasinio
(PNJL) model \cite{pnjl,liuyx}, Dyson-Schwinger equation \cite{dse},
statistical model \cite{yang,peter}, and the UrQMD and AMPT
transport models \cite{luo,zhim} etc. Each model has its merits and
results, but the consistency is lacking and the contradiction is
existing among the theoretical models. The difficulty is that the
fluid dynamical development of energy, momentum, and baryon charge
density leads to a complex spatial distribution of the measured and
statistically analyzed quantities at the freeze-out and
hadronization domain of the space-time. These also influence the
measured higher moments even without a phase transition, and to
disentangle the two effects is not easy \cite{wang}. Thus the
question is still open and further studies are required.

In this paper a new method is studied to provide more insight.
Namely, non-statistical moments of net proton event distribution are
calculated as the difference between the moments derived from real
events generated by the parton and hadron cascade model PACIAE
\cite{yan} and the ones from mixed events which are randomly
constructed according to the real events \cite{afan}. In this way the single
particle distribution arising from the fluid dynamic development is
separated from the two and more particle correlations stemming from
the PACIAE model where the hadronization and freeze-out takes place,
according to our expectation.

\begin{figure}
\centering
\includegraphics[width=2.5in]{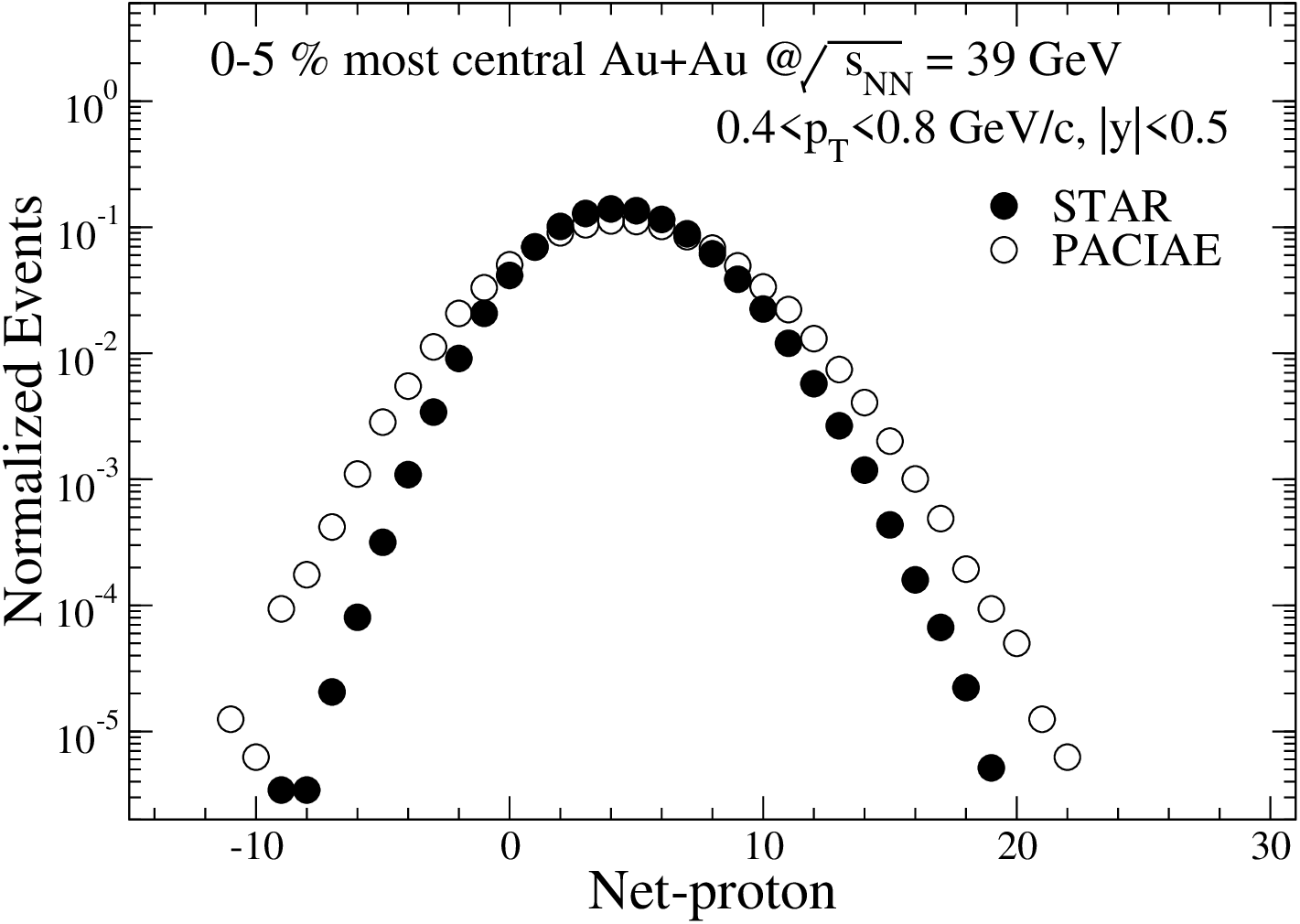}
\caption{
Net-proton event distributions in Au+Au collisions at $\sqrt{s_{NN}}$=39 GeV.
The solid and open symbols are STAR data (taken from first quotation of Ref.
\cite{luo}) and the results of PACIAE real events, respectively.}
\label{dis39_05}
\end{figure}

%%%%%%%%%%%%%%%%%%%%%%%%%%%%%%%%%%%%%%%
\section {Models}
%%%%%%%%%%%%%%%%%%%%%%%%%%%%%%%%%%%%%%%
The PACIAE model \cite{yan} is a parton and hadron cascade model,
which is based on PYTHIA \cite{PYTHIA}. PACIAE consists of the
parton initiation, parton evolution (rescattering), hadronization,
and the hadron evolution (rescattering) four stages.

In the parton initiation stage, a nucleus-nucleus collision is decomposed
into binary nucleon-nucleon (NN) collisions according to the collision
geometry and total NN cross section. The collision time is calculated for
each NN collision pair assuming straight line trajectory between two
consecutive NN collisions and the NN collision list is then constructed by
all collision pairs. A NN collision with earliest collision time is selected
from the collision list and performed by PYTHIA with string fragmentation
switches-off and diquarks (anti-diquarks) break into quark pairs (anti-quark
pairs). Thus a parton initial state (quarks, antiquarks, and gluons) is
eventually obtained when NN collision pairs are exhausted.

The parton rescattering is then proceeded by Monte Carlo method
using $2\rightarrow2$ leading order perturbative QCD cross sections
\cite{BL1977}. The parton evolution stage is followed by the
hadronization at the moment of partonic freeze-out (exhausting the
partonic collisions). The Lund string fragmentation model and a
phenomenological coalescence model are provided for hadronization.
After this the rescattering among produced hadrons is dealt with the
usual two body collision model \cite{yan}. Only the rescatterings
among $\pi$, $K$, $p$, $n$, $\rho (\omega) $, $\Delta$, $\Lambda$,
$\Sigma$, $\Xi$, $\Omega$, $J/\Psi$ and their antiparticles are
considered for simplicity.

Like other transport (cascade) models, such as above mentioned UrQMD
\cite{bass} and/or AMPT \cite{lin}, PACIAE does not assume
equilibrium. It just simulates dynamically the whole relativistic
heavy ion collision process from the initial partonic stage to the
hadronic final state via the parton evolution (not implemented in
UrQMD), hadronization, and hadron evolution according to a copious
dynamical ingredients (assumptions) introduced reasonably. Therefore
it is parallel to the experimental nucleus-nucleus collisions. These
dynamics correctly describes the particle, energy, and entropy etc.
developments, while intensive thermodynamical quantities are not
defined in this non-equilibrium regime. Messages brought by the
produced particles in these transport (cascade) models are all of
dynamical origin. Unlike most of the hydrodynamic models where the
phase transition is described via the assumptions for equation of
states, in PACIAE (the same in UrQMD and/or AMPT) we do not
implement a phase transition congenitally. However, once there is a
phase transition signal in transport (cascade) model calculations,
it must be a result of dynamical evolution. Of course, further
studies are then required.

The n$^{th}$ moment about the mean of a conserved quantity, $x$,
with the event distribution $P(x)$ is expressed as
\begin{equation}
M^{(n)}=\langle(x-\langle x \rangle)^n\rangle=\int{(x-\langle x
\rangle)^nP(x)dx}, \label{mea}
\end{equation}
where the n$^{th}$ moment about zero reads as
\begin{equation}
\langle x^{(n)}\rangle=\int{x^nP(x)dx},
\end{equation}
The higher moments as well as their products investigated widely are
then
%\begin{flushleft}
\begin{equation}
{\rm variance:} \hspace{1cm} \sigma^2=M^{(2)}, \label{sig}
\end{equation}
\begin{equation}
{\rm skewness:} \hspace{1cm} S=M^{(3)}/(M^{(2)})^{3/2}, \label{ske}
\end{equation}
\begin{equation}
{\rm kurtosis:} \hspace{1cm} \kappa=M^{(4)}/(M^{(2)})^2-3,
\label{kur}
\end{equation}
%\end{flushleft}
\begin{equation}
S\sigma=M^{(3)}/M^{(2)},
\label{ssig}
\end{equation}
and
\begin{equation}
\kappa\sigma^2=M^{(4)}/M^{(2)}-3M^{(2)},
\label{ksig}
\end{equation}
besides the mean $M\equiv\langle x^{(1)}\rangle$ and $M^{(1)}\equiv 0$.

\begin{figure}
\centering
\includegraphics[width=3.4in,clip=true]{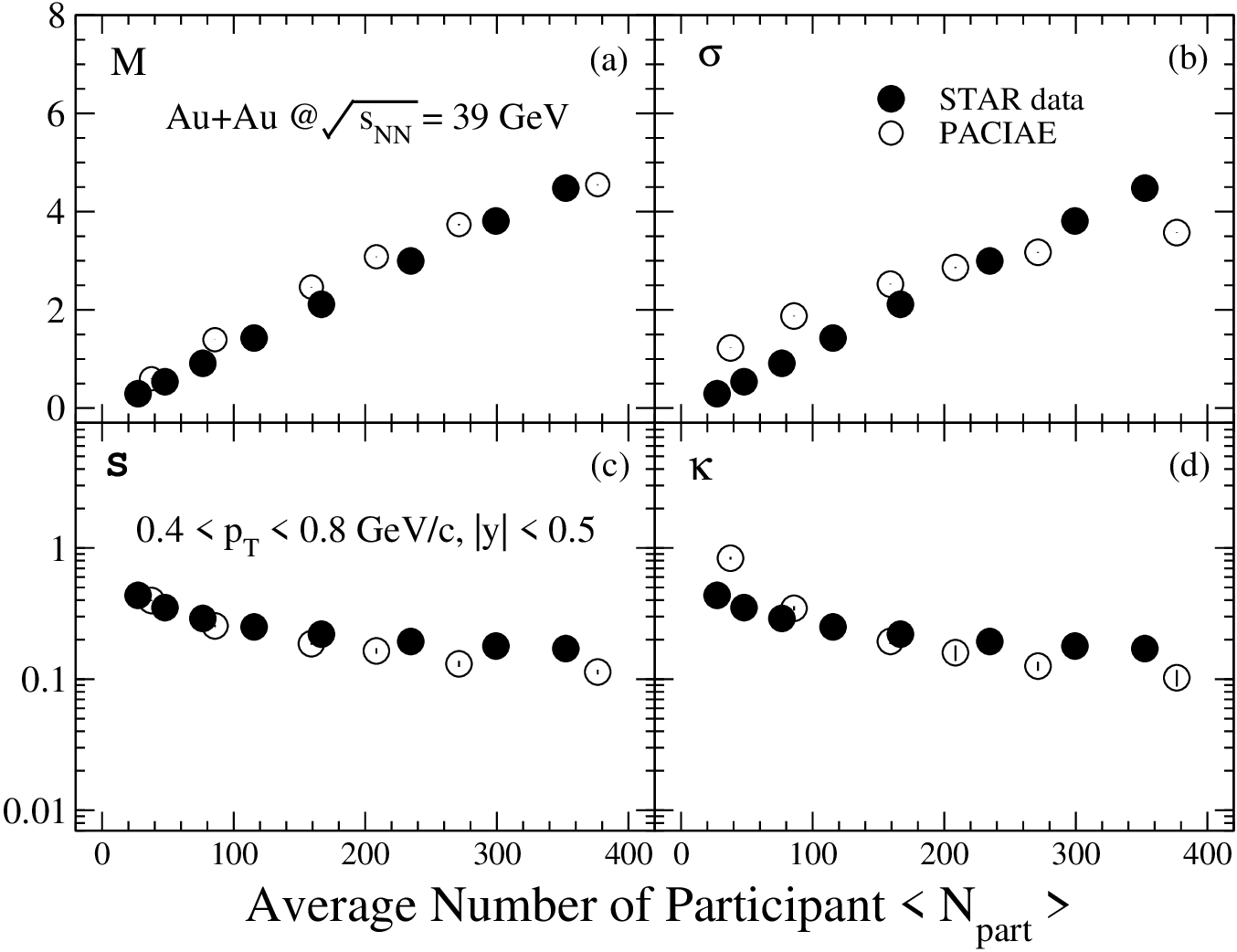}
\caption{
Centrality dependent moments of net-proton event distribution in Au+Au
collisions at $\sqrt{s_{NN}}$=39 GeV. The solid and open circles are STAR
data (taken from first quotation of Ref. \cite{luo}) and the results of PACIAE
real events, respectively. }
\label{compau39}
\end{figure}

The study of higher moment singularities is a matter of dynamics, but the
statistical fluctuation is always dominant. Therefore, we study the
non-statistical part of fluctuations instead of full fluctuations in this
paper. The parton and hadron cascade model PACIAE \cite{yan} is applied to
generate real events. Each real event obeys dynamical conservation laws (such
as net baryon number conservation, energy and momentum conservations, etc.),
which cause correlations among particles in a single event. As the PACIAE
model simulation for a nucleus-nucleus collision is parallel to the
experiment of a nucleus-nucleus collision, the NA49 method \cite{afan} is
also employed to generate the mixed events. This means that the mixed events
are constructed by combining particles randomly selected from different real
events, while reproducing the event multiplicity distribution of real events.
We have checked that the dynamical conservation laws really do not exist in
the mixed event and there are only statistical fluctuations caused by the
effects of finite event number, finite size, and the experimental finite
detector resolutions. The ``non-statistical moments" of conserved quantity
event distributions are defined to be the difference between the moments
derived from real events and the ones from mixed events. Thus the dynamical
fluctuations are pronounced from the underlying dominant statistical
fluctuations and are expected to be seen easily in the non-statistical higher
moments.

If the n$^{th}$ moment about the mean calculated by real (mixed) events is
denoted by $M^{(n)}_R$ ($M^{(n)}_M$), then the corresponding n$^{th}$
non-statistical moment about the mean, like in \cite{afan}, is:
\begin{equation}
M^{(n)}_{NON}=M^{(n)}_R-M^{(n)}_M.
\label{rm}
\end{equation}

\begin{figure}
\centering
\includegraphics[width=2.4in,clip=true]{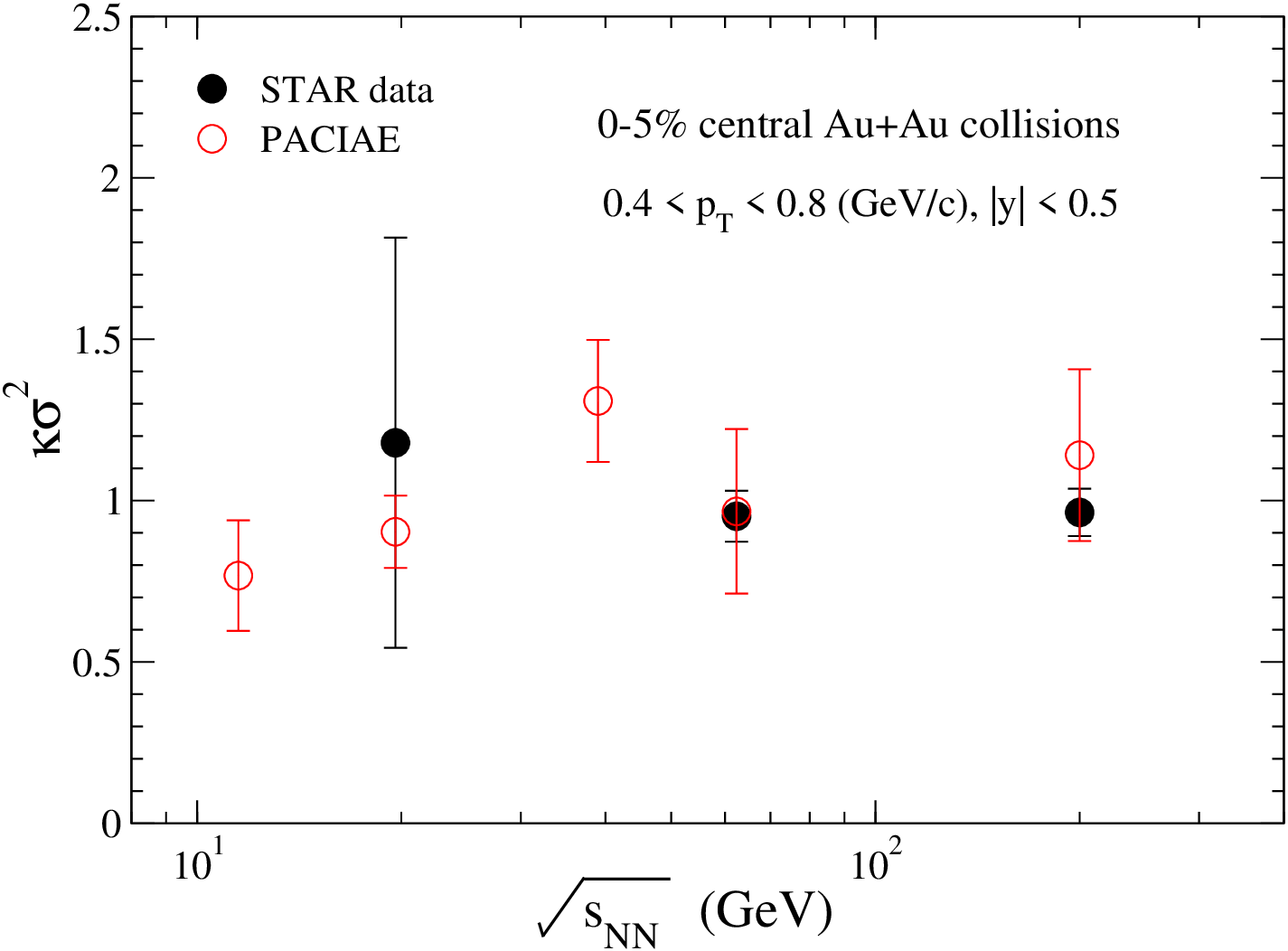}
\caption{(Color online) Energy dependence of $\kappa\sigma^2$ of net-proton
event distribution in Au+Au collisions. The solid and open circles are STAR
data (taken from \cite{star}) and the results of PACIAE real events,
respectively.}
\label{kursgcomp}
\end{figure}

%%%%%%%%%%%%%%%%%%%%%%%%%%%%%%%%%%%%%%%
\section {Results}
%%%%%%%%%%%%%%%%%%%%%%%%%%%%%%%%%%%%%%%
The net-proton event distribution and the corresponding higher moments in
relativistic Au+Au collisions at $\sqrt{s_{NN}}$ from 11.5 to 200 GeV are
calculated with both the PACIAE real events and the corresponding mixed
events. It is found that the PACIAE results of mean, variance, skewness, and
kurtosis in Au+Au collisions are in agreement with STAR data.

Shown, as an example, in Fig.~\ref{dis39_05} are the results of PACIAE real
events (open symbols) compared with the STAR data (solid symbols) of
net-proton event distribution in 0-5\% most central Au+Au collision at
$\sqrt{s_{NN}}$=39 GeV. Note that the spectator protons have been excluded
here. One sees from Fig.~\ref{dis39_05} that the agreement between the STAR
data and PACIAE results is satisfactory: with two distributions having nearly
the same peak location and close each other until the half height. The
deviation of the two distributions gets visible around the tails, however,
the contribution of the tails to the moments is expected very small as the
corresponding probability is rather low. In our knowledge there was not yet
comparison between the STAR data of net-proton event distribution and model
calculations, except the parameter fit in \cite{yang}.

The centrality dependent STAR data of the moments of net-proton event
distribution in Au+Au collision at $\sqrt{s_{NN}}$=39 GeV (cf. first quotation
of Ref. \cite{luo}) are compared with the results of PACIAE real events in
Fig.~\ref {compau39}. In this figure one sees the agreement between the STAR
data and PACIAE results for $M$ and $S$. However the agreement in $\sigma$ and
$\kappa$ is not as good as in $M$ and $S$. This reveals that the tails of the
net-proton event distribution affect the even moments stronger.

Fig.~\ref{kursgcomp} shows the energy dependence of the STAR $\kappa\sigma^2$
data and the results of PACIAE real events in Au+Au collisions. The STAR
data characteristic feature of almost energy independent $\kappa\sigma^2$ is
approximately reproduced by PACIAE within error bars.

\begin{figure}
\centering
\includegraphics[width=3.5in]{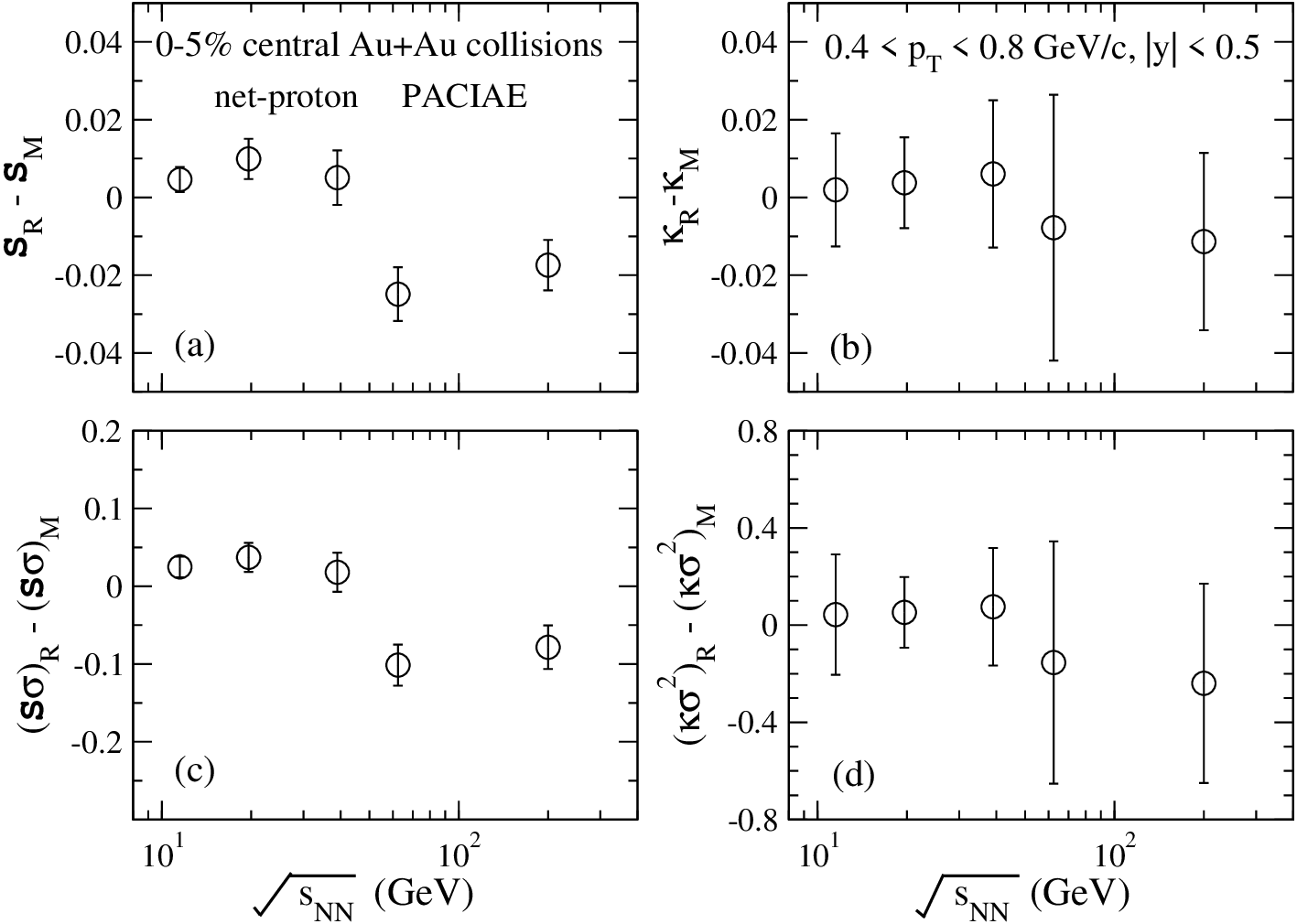}
\caption{Energy dependence of non-statistical higher moments and their
products of the net-proton event distribution in 0-5\% most central Au+Au
collisions.} \label{real-mixed}
\end{figure}

We give in Fig.~\ref{real-mixed} the PACIAE results of energy dependence of
the non-statistical skewness (panel a), kutorsis (b), $S\sigma$ (c), and
$\kappa\sigma^2$ (d) calculated for the 0-5\% most central Au+Au collisions
in the STAR acceptances of $|y|<0.5$ and $0.4<p_T<0.8$ GeV/c according to
Eqs.~(\ref{mea}-\ref{rm}). The energies applied in the calculations are
$\sqrt{s_{NN}}$=11.5, 19.6, 39, 62.4 and 200 GeV, with the total number of
generated real events being 3.0 $\times 10^5$, 3.0 $\times 10^5$, 2.4 $\times
10^5$, 1.2$\times 10^5 $ and 1.2 $\times 10^5$, respectively. The same number
of mixed events are then constructed correspondingly. It is found in this
figure that the non-statistical $S$, $\kappa$, $S\sigma$, and $\kappa\sigma^2$
change sign at $\sqrt{s_{NN}}\sim$ 60 GeV. Here a first evidence of
singularity is seen in the energy dependent $S$ and $S\sigma$. These signs,
of course, are not seen in the results calculated by mixed events. They are
also not showing up in the results calculated by real events and in the STAR
data (results of full fluctuations indeed) \cite{gupta,luo1}, because
dynamical fluctuations are always submerged in the statistical fluctuations.
It seems that the higher moment singularity be implicated in the
non-statistical fluctuation.
\vspace{2cm}

\begin{figure}
\centering
\includegraphics[width=3.5in,clip=true]{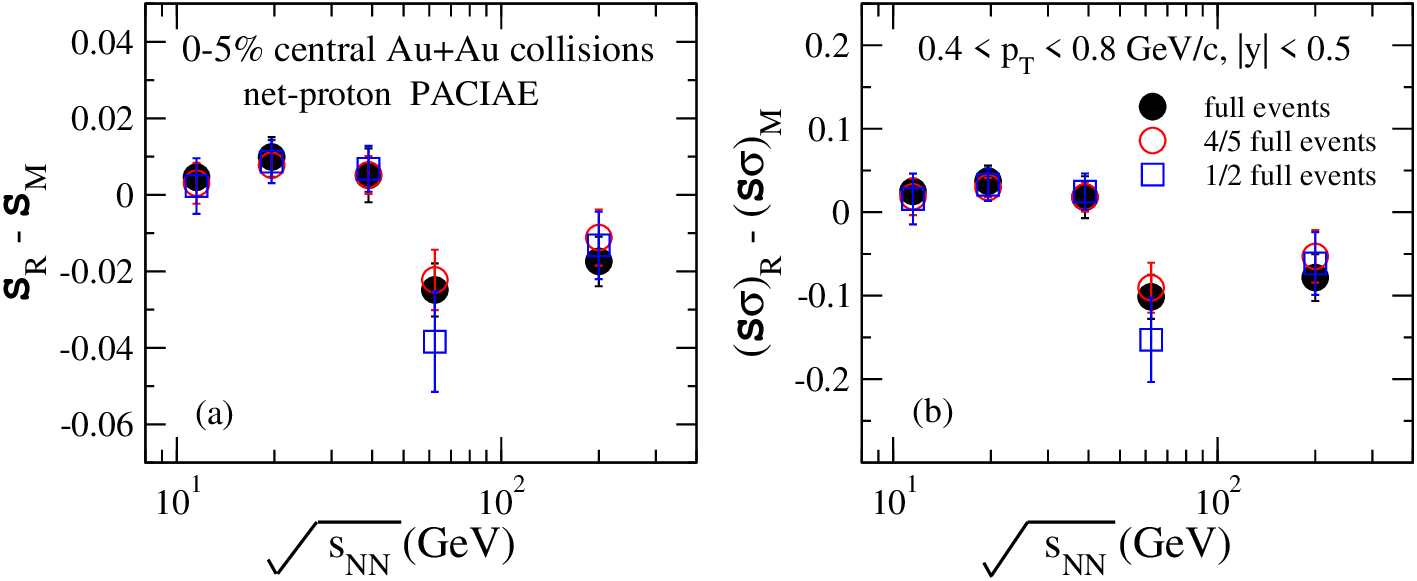}
\caption{(Color online) Energy dependence of non-statistical $S$ and
$S\sigma$ of the net-proton event distribution in 0-5\% most central Au+Au
collisions.}
\label{dfevent}
\end{figure}

To check the reliability of the sign changes in
Fig.~\ref{real-mixed}, we have recalculated the energy dependent
non-statistical $S$ and $S\sigma$ with $1/2$ and $4/5$ total number
of events and compared the results with the ones calculated with the
full number of events, as shown in Fig~\ref{dfevent}. We see in this
figure that the convergence is quite good for $\sqrt{s_{NN}}$= 11.5,
19.6, 39 and 200 GeV and satisfactory for 62.4 GeV. Thus the results
shown in Fig.~\ref{real-mixed} are reliable. The sign changes of $S$
and $S\sigma$ shown in this figure may be a response to the
prediction in \cite{asaka}.

The results in Fig.~\ref{real-mixed}, when compared to the estimates
in the second quotation of \cite{cser}, indicate that the freeze-out
point is very close to the hadronization point as the value of
skewness is rather small. Furthermore, at lower beam energies the
skewness is positive, indicating a freeze-out closer to the QGP side
in the hadronization process. But at higher beam energies the
skewness is negative, indicating more dominance of the hadronic side
of the phase transition. The kurtosis becomes significantly
negative, although small at higher beam energies, indicating that
the freeze-out is indeed in the phase transition domain. Since a
number of dynamical ingredients are introduced in the PACIAE
(PYTHIA) model, the detailed roles and effects of the dynamical
ingredients relevant to this phenomenon have to be studied later.

%%%%%%%%%%%%%%%%%%%%%%%%%%%%%%%%%%%%%%%
\section {Conclusion}
%%%%%%%%%%%%%%%%%%%%%%%%%%%%%%%%%%%%%%%
In summary, we have calculated the higher moments of net-proton event
distributions in relativistic Au+Au collisions at $\sqrt{s_{NN}}$ from 11.5
to 200 GeV with real events generated by the parton and hadron cascade model
PACIAE \cite{yan}. The PACIAE results of centrality dependent net-proton $M$,
$\sigma$, $S$, and $\kappa$ in Au+Au collisions are consistent with the STAR
data.

We have studied the non-statistical fluctuations of net-proton event
distributions. The mixed events are constructed according to PACIAE real
events and the non-statistical moments are calculated as the difference
between the moment calculated from real events and the one from mixed events.
The non-statistical $S$, $\kappa$, $S\sigma$, and $\kappa\sigma^2$ appear to
change signs at $\sqrt{s_{NN}}\sim$ 60 GeV. The clear sign change in the
energy dependent non-statistical $S$ and $S\sigma$ may reflect some
singularities. However, this has to be confirmed by the STAR Beam Energy Scan
non-statistical fluctuation data later.

Acknowledgements: This work was supported by the National Natural Science
Foundation of China under grant nos.: 10975062, 11075217, 11105227, 11175070,
and by the 111 project of the foreign expert bureau of China. AL and YPY
acknowledge the financial support from TRF-CHE-RMUTI under contract No.
MRG5480186. Authors thank X. F. Luo for the STAR data. We are grateful to
N. Xu for the valuable discussions.

\end{document}